%Paper: cond-mat/9211016
%From: jysoo@hlrserv.hlrz.kfa-juelich.de (Jysoo Lee)
%Date: Mon, 23 Nov 92 11:46:43 +0100

\magnification=1200
\baselineskip=20pt
\normallineskip=10pt
\overfullrule=0pt
\footline={\sevenbf Lee/Herrmann --- July 21, 1992 \hss PAGE \folio}

\null
\vfill
\centerline {\bf Angle of Repose and Angle of Marginal Stability:}
\medskip

\centerline {\bf Molecular Dynamics of Granular Particles}
\bigskip
\bigskip

\centerline {Jysoo Lee and Hans J. Herrmann}
\medskip

\centerline {HLRZ-KFA J\"{u}lich, Postfach 1913, W-5170 J\"{u}lich, Germany}
\bigskip
\vfill

\centerline{\bf Abstract}
\bigskip

We present an implementation of realistic static friction in molecular
dynamics (MD) simulations of granular particles. In our model, to
break contacts between two particles, one has to apply a finite amount
of force, determined by the Coulomb criterion. Using a two dimensional
model, we show that piles generated by avalanches have a {\it finite}
angle of repose $\theta_R$ (finite slopes). Furthermore, these piles
are stable under tilting by an angle smaller than a non-zero tilting
angle $\theta_T$, showing that $\theta_R$ is different from the angle
of marginal stability $\theta_{MS}$, which is the maximum angle of
stable piles. These measured angles are compared to a theoretical
approximation. We also measure $\theta_{MS}$ by continuously adding
particles on the top of a stable pile.
\bigskip


\vfill
\eject
\null

\noindent
{\bf 1. Introduction}
\bigskip

Systems of granular particles (e.g. sands) exhibit many interesting
phenomena.$^{1-3}$ The formation of spontaneous heap$^{4-6}$ and
convection cells $^{7-11}$ under vibration, density waves found in the
outflow through hoppers$^{12-16}$ and segregation of
particles$^{17-20}$ are just a few examples. These phenomena are
consequences of the unusual dynamical response of the system. One of
the characteristic property of granular systems is that it can behave
like both a solid and a fluid. One can pour (like a fluid) sand grains
on a table, and they form a stable pile with finite slope (like a
solid). Part of the reason why it acts like a solid is due to static
friction. By static friction, we mean that one has to apply a force
larger than certain {\it threshold} in order to break a contact
between particles. This threshold is determined by, for example, the
Coulomb criterion. Static friction is responsible for many static
structures (e.g. sand pile), and have a possible implication in the
dynamics of granular systems.$^{21}$ Despite its importance, the
effect of static friction has been much less studied as compared to
other microscopic mechanisms. This is mainly due to the difficulty of
including static friction to a theoretical framework or a simulational
scheme.
\medskip

In this paper, we present an implementation of static friction in a
molecular dynamics (MD) simulation, which uses a scheme introduced by
Cundall and Strack.$^{22}$ Using this code, we generate piles by first
filling a (two dimensional) box with grains, then removing a sidewall.
The slope of the pile is finite, which is related to the finite
``angle of repose ($\theta_R$).'' Here, $\tan \theta_R$ is defined to
be the slope of the pile. This angle is strongly dependent on the
friction coefficient $\mu$, and rather insensitive to other parameters
of the system. Furthermore, we find that the pile obtained above is
stable under tilting by an angle smaller than the finite tilting angle
$\theta_T$, where $\theta_T$ is typically a few degrees. This suggests
that the angle of marginal stability $\theta_{MS}$, the maximum angle
of stable piles, is {\it larger} than $\theta_R$, which has been
observed for real sandpile experiments.$^{3,23}$ We also study the
situation of keeping on adding particles on a stable pile. The angle,
at which the pile becomes unstable, can be interpreted as
$\theta_{MS}$. We also propose a theoretical method to calculate
$\theta_{MS}$ and $\theta_T$ from an approximate stress distribution
obtained by Liffman {\it et al}.$^{24}$ The theoretical values show
similar dependency of the measured angles on the friction coefficient
$\mu$.
\bigskip
\bigskip

\noindent
{\bf 2. Definition of the model}
\bigskip

The interaction between real sand grains is too complicated to
construct a model, by which {\it all} the properties of a granular
system are described accurately. Instead of constructing a model to
reproduce all the details from the beginning, it is often advantageous
if one identifies the basic ingredients of the system, construct a
model with these ingredients. It is often true that the qualitative
behavior of a system is independent of the fine details of the model.
Some important ingredients for a granular system are (1) repulsion
between two particles in contact, (2) dissipation of energy during
collision. In certain cases, the rotation of the particles could be
important.$^{25}$ In the previous molecular dynamics (MD) simulations
of granular systems, most of these ingredients, if not all, are
incorporated. For example, the repulsion and dissipation is included
in most MD simulations of granular particles.$^{10,11,16,18,22,24-27}$
A few of these simulations also included the rotational degree of
freedom.$^{16,18,22,24-26}$ Here, we will construct a model which
includes the repulsion, dissipation and static friction. But, the
model does not have rotational degrees of freedom.
\medskip

An individual grain is modeled by a spherical particle. These
particles interact with each other only if they are in contact.
Consider two particles $i$ and $j$ in contact in two dimenstion. Let
the coordinate for the center of particle $i$ ($j$) be $\vec {R}_i$
($\vec {R}_j$), and $\vec {r} = \vec {R}_i - \vec {R}_j$. A vector
$\vec {n}$ is defined to be a unit vector parallel to the line joining
the centers of two particles, $\vec {r}/ r$. Another vector $\vec
{s}$, which is orthogonal to $\vec {n}$, is obtained by rotating $\vec
{n}$ by $\pi / 2$ in clockwise direction. We also define the relative
velocity $\vec {v}$ to be $\vec {V}_i - \vec {V}_j$, and the radius of
particle $i$ to be $a_i$.
\medskip

The force on particle $i$ exerted by particle $j$, $\vec {F}_{j \to
i}$, can be written as
$$
\vec {F}_{j \to i} = F_n \vec {n} + F_s \vec {s}, \eqno (1)
$$
\noindent
where the normal force $F_n$ is given by
$$
F_n = k_n (a_i + a_j - r)^{3/2}- \gamma_n m_e (\vec {v}
       \cdot \vec {n}). \eqno (2a)
$$
\noindent
The first term of (2a) is the three-dimensional Hertzian repulsion due
to the elastic deformation, where $k_n$ is the elastic constant of the
material. The second term is a velocity dependent friction term, which
is introduced to dissipate energy from the system. Here, $\gamma_n$ is
a constant controlling the amount of dissipation, and $m_e$ is the
effective mass $m_i m_j / (m_i + m_j).$ The second term of (1), the
shear force $F_s$ is
$$
F_s = - \gamma_s m_e (\vec {v} \cdot \vec {s}) - {\rm sign} (\delta s)
\cdot {\rm min}(k_s \delta s, \mu F_n), \eqno (2b)
$$
\noindent
where the first term is a velocity dependent damping term similar to
the one in (2a). The second term of (2b) simulates the static
friction. The basic idea is the following.$^{22}$ When two particles
start to touch each other, one puts a ``virtual'' spring in the shear
direction. For the {\it total} shear displacement $\delta s$ during
the contact, there is a restoring force, $k_s \delta s$, which is a
counter-acting frictional force. The maximum value of this restoring
force is given by Coulomb's criterion---$\mu F_n$. When particles
are no longer in contact with each other, we remove the spring. We
want to emphasize that {\it one has to know the total shear
displacement of particles during the contact, not the instantaneous
displacement} to calculate the static friction. In other words, one
has to remember whether a contact is new or old. The rotation of the
particles is not included in the present simulation.
\medskip

The particles can also interact with walls. If particle $i$ is in
contact with a wall, the force exerted by the wall on the particle has
exactly the same form as eqs. (2) with $a_j = 0$ and $m_j =
\infty$. Also, there is a gravitational field. The force acting on
particle $i$ by the field is $-m_i g$. The total force acting on
particle $i$ is the vector sum of the particle-particle
interaction(s), the particle-wall interaction(s) and the gravitational
force.
\medskip

The trajectory of a particle is calculated by the fifth order
predictor-corrector method.$^{28}$ We use two Verlet tables. One is a
usual table with finite skin. The other table is a list of pairs of
{\it actually} interacting particles, which is used to calculate the
static friction term. For a typical situation, the CPU time needed to
run $1372$ particles is about $0.01$ seconds per iteration on a
Cray-YMP, which is comparable to the speed of the layered-link-cell
implementation of a short range Lennard-Jones system.$^{29,30}$
\bigskip
\bigskip

\noindent
{\bf 3. Obtaining the angle of repose}
\bigskip

As a non-trivial check whether the static friction term is working, we
measure $\theta_R$ as follows. We start by randomly putting $N$
particles in a box of width $W$ and height $H$. The particles fall
down due to gravity, and loose their energy due to dissipation. After
a long time, they fill the box with no significant motion. We show, in
Fig.~1(a), an example of the system at this stage. Here, parameters
are $\mu = 0.2, k_n = 10^6, k_s = 10^4, \gamma_n = 5 \times 10^2,
\gamma_s = \gamma_n/100$, and for walls, $k_n$ is chosen to be $2
\times 10^6$. We checked the motions of the particles by monitoring
the total kinetic energy of the system.  The average kinetic energy
per particle during the whole sequence is shown in Fig.~1(b). The
kinetic energy sharply rises when the wall is removed. The pile
relaxes in an oscillatory manner (see Fig.~1(b)). Next, we remove the
right wall, let the particles move out of the box, and wait until the
system reaches a new equilibrium. The figure 1(c) is the equilibrium
reached by starting from Fig.~1(a). As shown in the figure, the new
state has non-zero slope.
\medskip

We try a few ways to measure $\theta_R$. We first divide the box into
several vertical cells, the width of each cell is equal to the average
diameter of the particles. For the center of each cell $x$, we find
the maximum position $h(x)$ of the particles in the cell. The line
joined by these positions is a ``surface'' of the structure. Having
determined $h(x)$, we use three different ways to measure the slope:
(1) By joining $h(0)$ and $h(W)$, (2) by fitting a straight line to
$h(x)$ using the method of least squares, and (3) by fitting a
parabola to $\sum _{i=0}^{x} h(i)$ by the least squares method. Here,
if h(i) is a straight line, the sum is a parabola. In the case of
$h(x)$ being a straight line, these three methods should give
identical results. In our simulation, the slopes obtained by different
methods differ from each other by a few degrees.  For example, we
obtain $(1)~20.14 \pm 2.15,~(2)~18.90 \pm 1.49$ and $(3)~17.88 \pm
1.76$ for $400$ particle system with $\mu$ = 0.2. Here the angles are
averaged over $10$ samples. We find, on the average, the angle by
method (1) is larger than (2), and (2) is larger than (3), although
they are within the error bars of each other. It is quite possible
that these differences come from the finite size of the system. From
now on, we use only the method 3 for calculating the slope.
\bigskip
\bigskip

\noindent
{\bf 4. Parameter dependences}
\bigskip

For a fixed set of parameters which specify all the interactions, we
study the dependence of $\theta_R$ on the geometry, namely the aspect
ratio ($H/W$) of the box and the linear size of the system.  The
parameters are $k_n = 10^6, k_s = 10^4, \gamma_n = 5 \times 10^2,
\gamma_s = \gamma_n/100$. For walls, $k_n$ is chosen to be $2 \times
10^6$ to prevent particles to escape from the box. Since a system of
particles with equal radii tends to form a hexagonal packing, we use
particles with different sizes. The radii of the particles are drawn
from a Gaussian distribution with the mean of $0.1$ and width of
$0.02$, and the maximum (minimum) cut-off radius of a particle is
$0.13$ ($0.07$). In Fig.~2(a), we show the dependence of the angle
$\theta_R$ on the height $H$, for values of $\mu = 0.2$ and the width
$W = 2.0$. Each angle is obtained by averaging over $20$ samples. The
error plotted in Fig.~2(a) is the mean square sample-to-sample
fluctuations. Here, we cannot see any systematic dependence on $H$.
Also for other values of $\mu$, we find that $\theta_R$ does not
depends on the aspect ratio, as long as the ratio is sufficiently
larger than the slope of the pile generated. We then fix the aspect
ratio to be $2$, and study the dependence on the size of the system.
The angle $\theta_R$ for different values of $W$ is shown in
Fig.~2(b). All angles, as well as those presented in Fig.~2(b), are
averaged over $20$ samples, unless specified otherwise. For $\mu =
0.2$, these angles decrease for small sizes, and seem to saturate
starting around $W = 3.5$. For larger values of $\mu$, the angle
saturates for larger values of $W$.  For example, the angle continues
to decrease until $W = 4.0$ for $\mu = 0.3$. On the other hand, for
smaller $\mu$, the angle saturates for smaller $W$.  For $\mu = 0.1$,
there is no obvious trend of the data, even up to the very small
values of $W = 1.5$.  Since we want the angle obtained by this
simulation not to suffer from a finite size effect, we will use in the
followings the values $W = 4$ and $H/W = 1$ to calculate $\theta_R$.
We will also study cases of $\mu$ not larger than $0.2$.  The
simulation for larger values of $\mu$ is limited due to the fact that
one needs a larger aspect ratio and system sizes to be free of any
finite size effects.
\medskip

Next, we study how $\theta_R$ depends on the various interaction
parameters in the system: $\gamma, k$ and $\mu$. In a static
configuration, the damping term is absent, so ideally $\gamma$ terms
do not change $\theta_R$. However, since we prepared the sandpile by a
dynamical method (by causing an avalanche), the angle may depend on
$\gamma$. Also, $k_s$ is an the elastic constant of the artificial
spring we introduced, so it should not make a difference in a static
configuration, as long as we keep the value in a reasonable range. The
quantity of particular interest is the friction coefficient $\mu$,
since $\mu$ determines if contacts between particles are stable
(``stick'') or unstable (``slip''). Since the stability of the whole
structure (e.g. a pile) will be strongly influenced by that of
individual contacts, we expect $\theta_R$ will be strongly dependent
on $\mu$. For example, $\theta_R$ should be zero for $\mu = 0$, if the
individual grains in the pile are not moving. We first study the
effect of $k_n$ and $\gamma_n$ on $\theta_R$. We will limit ourselves
only to study the general trend such as the direction and the order of
magnitude of the changes. We also fix $\mu = 0.2$. We measure the
angle (inside parenthesis) for three different values of
$k_n$,~$10^4~(16.53 \pm 0.73),~10^5~(18.65 \pm 0.70) $ and
$10^6~(17.99 \pm 0.64)$. The difference in angle is very small, and
there seems to be no systematic dependency. For three values of
$\gamma_n =~50,~100,~500$, the angles are $16.47 \pm 0.75,~17.15 \pm
0.62,~17.99 \pm 0.64$. The angle seems to decrease systematically, as
$\gamma_n$ is being decreased.  However, the magnitude of the changes
is still small ($\sim 5\%$).
\medskip

Now, we study the effect of $\mu$. In Fig.~3, we show $\theta_R$
obtained for several different values of $\mu$. In the range of $\mu$
we studied, there seems to be a linear relation between the angle and
$\mu$. This relation can be true for small values of $\mu$, but it
{\it can not} be true for the entire ranges of $\mu$. The maximum
$\theta_R$ is limited to $90$ degrees, while the value of $\mu$ can be
arbitrarily large. We will come back to discuss this relation later.
Note that there are two friction coefficients in the system, one
between particles and another between particles and walls. We will
argue that, for a sufficiently large pile, the friction coefficient
which determines $\theta_R$ is that between the particles. Consider a
sandpile on a table. The stress distribution near the top part of the
pile would not be altered by the stress distribution at the bottom of
the pile. Therefore, only the friction coefficient between particles
can change $\theta_R$ in this region. On the other hand, the stress
distribution near the bottom of the pile will greatly be influenced by
the particle-wall friction coefficient. So, we expect the angle be
different near the bottom of the pile, if the friction coefficients
are different from each other.
\medskip

The piles discussed above are generated by causing avalanches. In
experiments, the structure just after an avalanche (e.g. Fig.~1(c)) is
not critical, but stable. In other words, one must apply an additional
{\it finite force} to make the structure unstable. One way to apply
the force is by tilting the box which contains the pile. The tilting
angle $\theta_T$ is defined as the rotation angle at which the pile
becomes unstable. We want to emphasize that $\theta_T$ is shown to be
non-zero for real sandpile experiments. We measure the tilting angle
for our model as follows. Starting from the pile like one in
Fig.~1(c), we rotate clockwise the whole box with a constant rate of
$10^{-3}$ degree/iteration. Then, we record the angle at which the
pile starts to move, which is defined as the tilting angle $\theta_T$.
The tilting angle for several values of $\mu$ is shown in Fig.~4.
Here, the width of the box $W$ is $4$, and the aspect ratio is $1$.
Indeed, one needs a {\it finite} tilting angle for the piles generated
using our model, and it gives us confidence that the model studied
here is reproducing the behavior of realistic static friction. The
finite $\theta_T$ implies that the pile is stable (not critical),
therefore $\theta_{MS}$ should be larger than $\theta_R$.
\bigskip
\bigskip

\noindent
{\bf 5. Pile with constant flux}
\bigskip

In the previous section, we argued that $\theta_R$ is smaller than
$\theta_{MS}$ based on the fact that the pile is stable even if it is
tilted by a finite angle smaller than $\theta_T$. We now propose a
method of obtaining the angle of marginal stability as well as the
angle of repose. Consider an empty box without a right wall, and put
one layer of particles at the bottom.  We monitor the maximum velocity
of the particles. If the maximum velocity is smaller than certain
value $v_{cut}$, then we insert a new particle at the upper left
corner of the pile. Once the particle is added, we then wait until the
maximum velocity of particles is again smaller than $v_{cut}$, then
add a new particle. This procedure is repeated long time for good
statistics.  In Fig.~5, we show the angle of the pile just before one
inserts a new particle. Here, $\mu = 0.2$, $W = H = 4.0$ and $v_{cut}
= 0.1$. We also simulate the system with $v_{cut} = 0.01, 0.001$, and
find no essential difference. The angle is zero at the beginning, and
increasing until it seems to just fluctuate for iterations larger than
$4 \times 10^5$. The curve shown in the figure is quite noisy, which
suggests that many configurations (or packings) are possible in the
steady state. The maximum angle of the pile one can build up before
avalanches is larger than the $\theta_R$ obtained before. This could
be an additional evidence that our model reproduces the difference
between $\theta_{MS}$ and $\theta_R$. We can estimate the difference
to be of the order of distance between two dotted lines in Fig.~5.
Here, the dotted lines represents the mean square fluctuations of the
angle.
\bigskip
\bigskip

\noindent
{\bf 6. Theoretical approach}
\bigskip

In the previous section, we measured various angles $\theta_R$,
$\theta_T$ and $\theta_{MS}$ for the model sand. How can we understand
these angles? In order to calculate these angles, one should know the
stress field inside the pile. Liffman {\it et al}$^{24}$ suggested an
approximate way of calculating the field in a packing of equal sized
spheres, which is illustrated in Fig.~6. In order to calculate the
stress at a point $O$, one draws two lines of slope $\sqrt {3}$ (line
$OA$) and $-\sqrt {3}$ (line $OB$) starting from the point $O$. Then,
the length of the lines ($l_A$ and $l_B$) within the pile (and above
the point) is approximately proportional to the force exerted by the
pile. For a more detailed explanation as well as for the justification
of this procedure, see Ref.~26. From this stress field, we calculate
$\theta_{MS}$ as follows. Consider a point at the bottom of the pile.
The normal (to the bottom surface) force at that point is $(l_A + l_B)
\sin (\pi/3),$ while the tangential force is $(l_A - l_B)
\cos (\pi/3)$. We then apply the Coulomb criterion. If the ratio of
the tangential to the normal force is larger than $\mu$, the contact
is unstable. In this way, for given $\mu$, we obtain the range of
angles at which the pile is stable. The largest angle at which the
pile is stable is the angle of marginal stability, which is
$$
\theta_{MS} = \tan^{-1}(3 \mu ).			\eqno (3)
$$
\noindent
Since $\theta_R + \theta_T$ is approximately the angle of marginal
stability, we plot both measured $\theta_R + \theta_T$ and calculated
$\theta_{MS}$ by the above procedure in Fig.~7(a). The difference
between the two angles is either due to the fact $\theta_{MS} \ne
\theta_R + \theta_T$ or the error of the approximation. In fact, the
approximation (and $\pi/3$ angle) is derived from an ordered packing
of the particles with same radius. It is possible that the stress
field in a disordered packing is very different from that of an
ordered one. In that case, one needs a new approximation scheme to
calculate the stress field. We also calculate $\theta_T$ for given
angle of repose $\theta_R$ and $\mu$. It is given by
$$
\mu = {R \cos (\pi/3-\theta_T) - \cos (\pi/3+\theta_T) \over
       R \sin (\pi/3-\theta_T) + \sin (\pi/3+\theta_T) },
\eqno (4a)
$$
\noindent
where
$$
R = {\tan(\pi/3+\theta_T) + \tan (\theta_R) \over
\tan(\pi/3-\theta_T) - \tan (\theta_R)} \cdot {\cos (\pi/3 + \theta_T)
\over \cos(\pi/3 - \theta_T)}.
\eqno (4b)
$$

\noindent
The $\theta_T$ obtained by eq.~(4) as well as the measured tilting angle
are plotted in Fig.~7(b). One can also see that the difference between
the two is small. Unlike the difference in Fig.~7(a), these two angles
should coincide if the stress distribution is calculated correctly.
\medskip

\noindent
The main point of presenting the theoretical approach is to show a
``first'' approximation for the problem, not to do quantitative
comparison between the measured and calculated angle. In order to
obtain more accurate numbers, one has to know a better way of
calculating the stress field inside the pile. However, it is
encouraging to see that even the values obtained by the first
approximation are comparable to the measured ones.
\bigskip
\bigskip

\noindent
{\bf Acknowledgments}
\bigskip
We thanks for many informative discussions with G. Ristow.
\vfill
\eject

\null
\noindent
{\bf References}
\bigskip

\item{1.} S. B. Savage, Adv. Appl. Mech. {\bf 24}, 289 (1984); S. B.
Savage, {\it Disorder and Granular Media} ed. D. Bideau,
North-Holland, Amsterdam (1992).
\medskip

\item{2.} C. S. Campbell, Annu. Rev. Fluid Mech. {\bf 22}, 57 (1990).
\medskip

\item{3.} H. M. Jaeger and S. R. Nagel, Science {\bf 255}, 1523
(1992).
\medskip

\item{4.} P. Evesque and J. Rajchenbach, Phys. Rev. Lett. {\bf 62}, 44
(1989).
\medskip

\item{5.} C. Laroche, S. Douady and S. Fauve, J. Phys. (France) {\bf
50}, 699 (1989).
\medskip

\item{6.} E. Clement, J. Duran and J. Rajchenbach, preprint.
\medskip

\item{7.} G. R\'{a}tkai, Powder Technol. {\bf 15}, 187 (1976).
\medskip

\item{8.} S. B. Savage, J. Fluid Mech. {\bf 194}, 457 (1988).
\medskip

\item{9.} O. Zik and J. Stavans, Europhys. Lett. {\bf 16}, 255 (1991)
\medskip

\item{10.} J. A. C. Gallas, H. J. Herrmann and S. Soko\l owski, Phys.
Rev. Lett., in press.
\medskip

\item{11.} Y-h. Taguchi, preprint.
\medskip

\item{12.} J. O. Cutress and R. F. Pulfer, Powder Technol. {\bf 1},
213 (1967).
\medskip

\item{13.} E. B. Pittman and D. G. Schaeffer, Comm. Pure Appl. Math.
{\bf 40}, 421 (1987).
\medskip

\item{14.} G. W. Baxter and R. P. Behringer, T. Fagaert and G. A.
Johnson, Phys. Rev. Lett. {\bf 62}, 2825 (1989).
\medskip

\item{15.} H. Caram and D. C. Hong, Phys. Rev. Lett. {\bf 67}, 828
(1991).
\medskip

\item{16.} G. Ristow, J. Physique I, {\bf 2}, 649 (1992).

\medskip

\item{17.} J. C. Williams, Powder Technol. {\bf 15}, 245 (1976).
\medskip

\item{18.} P. K. Haff and B. T. Werner, Powder Technol. {\bf 48}, 239
(1986).
\medskip

\item{19.} A. Rosato, K. J. Strandburg, F. Prinz and R. H. Swendsen,
Phys. Rev. Lett. {\bf 49}, 59 (1987).
\medskip

\item{20.} P. Devillard, J. Phys. (France) {\bf 51}, 369 (1990).
\medskip

\item{21.} For example, it is argued that the density waves formed in
Ref.~14 are due to ``arching,'' which is a consequence of static
friction. See also, ``bridge-collapsing'' in shear cells [Y. M. Bashir
and J. D. Goddard, J. Rheol. {\bf 35}, 849 (1991)].
\medskip

\item{22.} P. A. Cundall and O. D. L. Strack, G\'{e}otechnique {\bf
29}, 47 (1979).
\medskip

\item{23.} See, e.g., B. J. Briscoe, L. Pope and M. J. Adams, Powder Technol.
{\bf 37}, 169 (1984).
\medskip

\item{24.} K. Liffman, D. Y. C. Chan and B. D. Hughes, preprint.
\medskip

\item{25.} C. S. Campbell and C. E. Brennen, J. Fluid Mech. {\bf 151},
167 (1985).
\medskip

\item{26.} P. A. Thompson and G. S. Grest, Phys. Rev. Lett. {\bf 67},
1751 (1991).
\medskip

\item{27.} D. C. Hong and J. A. McLennan, preprint.
\medskip

\item{28.} D. J. Tildesley and M. P. Allen, {\it Computer Simulations
of Liquids}, Oxford University Press, Oxford (1987).
\medskip

\item{29.} G. S. Grest, B. D\"{u}nweg and K. Kremer, Comp. Phys. Comm.
{\bf 55}, 269 (1989).
\medskip

\item{30.} B. D\"{u}nweg, private communication.

\vfill
\eject

\null
\noindent
{\bf Figure Captions}
\bigskip

\item{Fig.~1:} (a) Box filled with $N = 1600$ particles just before
the right wall is removed. The thickness of the lines between centers
are proportional to the normal force. (b) Average kinetic energy per
particle (in erg) during the whole sequence of simulation. The energy
initially increases as particles fall down, then decays with time.
When the right wall is removed (iteration = $30000$), it increases
again. (c) Static pile obtained after the avalanche.
\medskip

\item{Fig.~2:} (a) The angle of repose $\theta_R$ vs the height $H$
with the width $W = 2$ and $\mu = 0.2$. The angle seems just
fluctuate, and no systematic dependency is found. (b) The angle of
repose $\theta_R$ vs $W$ for several values of $\mu$: $\mu = 0.1$
(diamond), $0.2$ (box) and $0.3$ (circle).  Here, the aspect ratio is
kept to be $2$.
\medskip

\item{Fig.~3:} The angle of repose $\theta_R$ vs $\mu$ with $W = H = 4$.
\medskip

\item{Fig.~4:} The tilting angle $\theta_T$ vs $\mu$ with $W = H = 4$.
\medskip

\item{Fig.~5:} The angle of pile measured when we add a new particle
to the pile. The two dotted lines indicate the width of the
fluctuation
\medskip

\item{Fig.~6:} The stress at a point inside the pile is approximately
the vector sum of line $A$ and $B$.
\medskip

\item{Fig.~7:} (a) The measured $\theta_R + \theta_T$ and the
calculated $\theta_{MS}$ is shown for several values of $\mu$. There
is a difference between the two. (b) The measured and calculated
$\theta_T$ for different values of $\mu$. The difference is smaller
than that of (a).
\vfill
\eject

\bye